\documentclass{bmvc2k}

\usepackage{times}
\usepackage{graphicx}
\usepackage{amsmath,amssymb,amsfonts,bm}
\usepackage{booktabs}
\usepackage{multirow}
\usepackage{algorithm}
\usepackage{algorithmic}
\usepackage{xcolor}
\usepackage{url}
\usepackage{hyperref}
\usepackage{siunitx}

\usepackage[percent]{overpic}
\usepackage{pict2e}
\usepackage{tikz}
\usetikzlibrary{arrows.meta}

\title{Lightweight Interpretable RGB-Guided Hyperspectral Super-Resolution under Real Cross-resolution Misalignment}

\addauthor{Mohamad Jouni}{https://mira-imaging.github.io/mj}{1}
\addauthor{Aurélien Godet}{aurelien.godet@grenoble-inp.fr}{1}
\addauthor{Mauro Dalla Mura}{mauro.dalla-mura@grenoble-inp.fr}{1,2}

\addinstitution{
	Univ. Grenoble Alpes, CNRS\\
	Grenoble INP, GIPSA-Lab\\
	38000 Grenoble, France
}

\addinstitution{
	Institut Universitaire de France (IUF)\\
	Paris, France
}

\runninghead{Jouni, Godet, Dalla Mura}{Confidence-Aware RGB-Guided HSR}

\begin{document}

\maketitle

\begin{abstract}
Compact snapshot hyperspectral cameras provide rich instantaneous spectral measurements for ground-level machine vision, but at lower spatial resolution than standard RGB cameras.
RGB-guided hyperspectral super-resolution (HSR) addresses this limitation by transferring spatial detail from a high-resolution RGB guide to a low-resolution hyperspectral image (HSI).
These dual-camera systems are typically in a horizontal rig geometry,
requiring cross-camera image alignment due to different fields of view. However, residual misregistration can inject spurious high-frequency details.
Existing learned unaligned-fusion methods are usually trained for a fixed spectral support and spatial scale factors and can be computationally demanding, limiting their flexibility across sensors.

We propose a lightweight and interpretable RGB-guided HSR framework combining cross-modal flow alignment with model-based Gram-Schmidt orthogonalization fusion.
The method first warps the RGB guide onto the HSI grid, then estimates an energy-based confidence weight map by measuring local alignment reliability.
This map is then used both in a weighted least-squares spectral regression and in a gated fusion between the RGB-guided SR estimate and an HSI-preserving estimate.
Unlike existing learned methods, the proposed framework has a low computational footprint and supports different VIS-NIR spectral supports and scale factors without retraining.

Experiments on the Real benchmark show that the proposed method improves reconstruction accuracy over learned fusion baselines while remaining substantially faster.
On a 34-frame sequence acquired with our real RGB-HSI dual-camera setup, a reduced-resolution quantitative evaluation validates the method under genuine cross-sensor radiometric, noise, and geometric differences, while native-resolution qualitative results demonstrate deployment on the full 51-band visible-near-infrared acquisition beyond the 31-band setting of existing learned baselines.
\end{abstract}

	
	\section{Introduction}
	\label{sec:intro}

	Hyperspectral imaging acquires dense spectral measurements across the electromagnetic spectrum, enabling material-level analysis beyond standard RGB imaging~\cite{manolakis2016hyperspectral}.
	Compact snapshot hyperspectral cameras are particularly attractive for ground-level machine vision because, unlike scan-based counterparts, they can instantly acquire spectral cubes, making them relevant for dynamic applications such as robotic inspection, precision agriculture, food analysis, non-destructive testing, heritage imaging, gas detection, and biomedical imaging~\cite{jouni2025multiple, 9050560, 10447567, contreras2025multimodal, ng2023hyper, schmitt2023rgb}.
	However, for miniaturized and low-cost systems, this additional spectral resolution comes at the cost of spatial resolution, especially for compact or snapshot hyperspectral cameras~\cite{hagen2013review}.
	A practical solution is therefore to combine a low-resolution (LR) hyperspectral image (HSI) with a high-resolution (HR) RGB image of the same scene, using the RGB image as a spatial guide for HSI super-resolution (HSR).
	This is a standard configuration in remote sensing~\cite{ciotola2024hyperspectral}, but in that case there is basically no parallax.

	\begin{figure*}[!t]
		\centering
		\includegraphics[width=\linewidth]{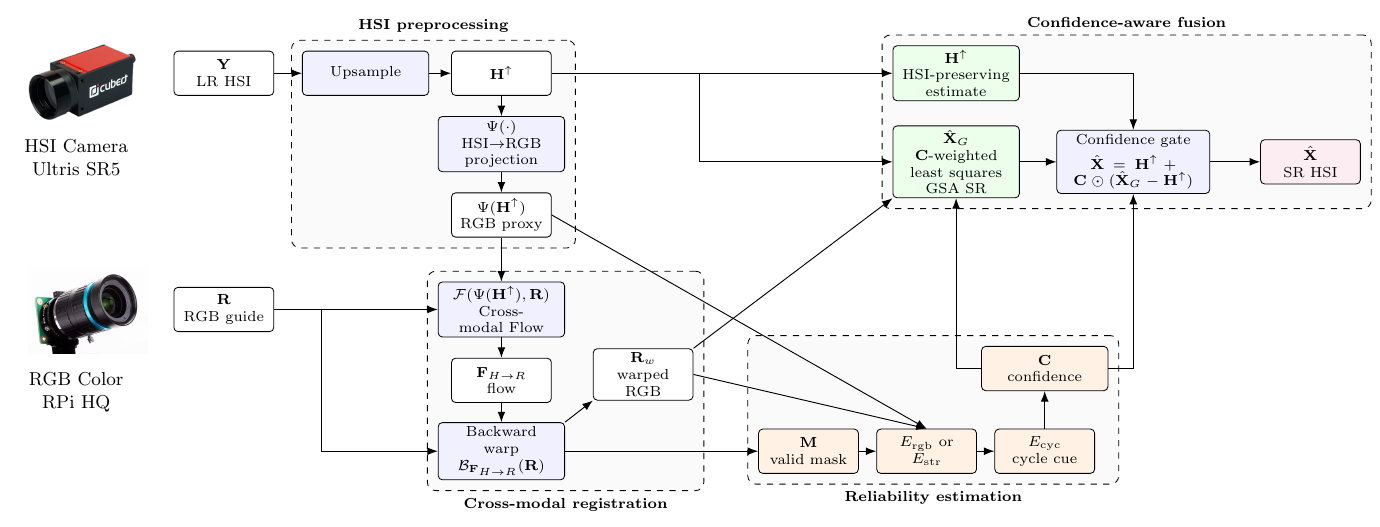}%
		\caption{Overview of the proposed confidence-aware RGB-guided HSI super-resolution framework, showing images of the Ultris SR5 HSI and the RPi HQ RGB cameras.}%
		\label{fig:pipeline}%
	\end{figure*}

	Most RGB-guided HSR methods assume that the RGB and HSI observations are accurately aligned.
	This assumption is difficult to satisfy in close-range dual-camera systems.
	Unlike far-field satellite settings, where viewpoint differences are often negligible at the scene scale, ground-level RGB-HSI rigs use separate sensors with distinct optics, a nonzero baseline, and different fields of view and spectral responses. Even after geometric calibration, the two cameras can observe different scene content near depth discontinuities, producing parallax and occlusions.
	The problem is further complicated by the cross-resolution nature of real RGB-HSI data. The HSI-derived RGB proxy is smoother and noisier than the RGB observation, which weakens dense correspondences around fine structures.
	
	Recent unaligned RGB-guided HSR methods address this problem with learned alignment and fusion modules~\cite{lai2025hsifn, zhang2025sschsr}.
	Such methods demonstrate the importance of registration-aware fusion, but they are computationally heavy, lack interpretability, and are tied to the spectral and spatial settings used during training. In particular, standard natural HSI benchmarks commonly use a 31-band visible-domain (VIS) protocol in the range $[400, 700]$~$\si{\nm}$~\cite{DeepCASSI:SIGA:2017, chakrabarti2011statistics, CAVE_0293, lai2025hsifn}, but in many applications it would be good to acquire information beyond this range~\cite{contreras2025multimodal, ng2023hyper, schmitt2023rgb}.
	Applying these models to another number of bands, another spectral range, or another scale factor generally requires retraining or spectral-domain adaptation~\cite{picone2026leveraging}.
	
	In contrast, lightweight model-based fusion methods offer a complementary trade-off.
	Gram-Schmidt adaptive (GSA) fusion~\cite{aiazzi2007improving} is time-efficient, interpretable, and uses multivariate spectral regression to adapt to different sensing configurations at a low computational footprint.
	However, such methods typically assume aligned inputs and do not handle locally unreliable registration.
	If a warped RGB guide is used without accounting for local registration failures, RGB details may be injected into the wrong HSI pixels, producing ghosting, edge duplication, and spectral distortion, especially with occluded and parallax areas.
	
	This paper investigates a complementary direction. Instead of replacing lightweight model-based fusion with a fully learned reconstruction network, we explicitly model when RGB guidance should be trusted. We build on a cross-modal flow (CMF) estimator \cite{zhou2022promoting} to align the RGB image to the HSI grid, and then compute a pixel-wise confidence map from registration validity and local inconsistencies. This confidence controls both the spectral regression used inside GSA and the final amount of detail injected into the HSI estimate. The resulting method is a reliability-aware version of CMF-guided GSA, designed for real RGB-HSI pairs with imperfect cross-resolution alignment.
	The novelty is not CMF or GSA individually, but their coupling through an explicit reliability map that controls both spectral regression and local detail injection.
	
	Our contributions are as follows. First, we formulate RGB-guided HSI SR under close-range cross-resolution misalignment as a confidence-aware registration-and-fusion problem. Second, we introduce an interpretable energy-based confidence map for warped RGB guidance, combining geometric validity with RGB-HSI consistency cues and bidirectional-flow consistency. Third, we extend GSA with confidence-weighted least-squares (WLS) spectral regression and Tikhonov regularization, reducing the influence of unreliable regions on spectral calibration. Finally, we formulate the reconstruction as confidence-gated detail injection between an HSI-preserving upsampled estimate and an RGB-guided GSA estimate.
	
	\section{Related Work}
	
	\paragraph{Hyperspectral image super-resolution.}
	Single-image super-resolution (SISR)~\cite{picone2026leveraging} reconstructs a HR HSI from a single LR HSI. ESSAformer~\cite{zhang2023essaformer}, for example, uses an efficient transformer to exploit spectral-spatial correlations.
	However, SISR remains limited by the lack of observed high-frequency spatial information.
	Fusion-based approaches instead use an auxiliary HR image, typically RGB, multispectral, or panchromatic data \cite{ciotola2024hyperspectral}.
	RGB-guided HSI SR has also been addressed through model-based optimization, such as TV$^3$ minimization~\cite{wan2022robust}.
	Model-based HSI fusion methods provide efficient detail injection, while modern deep networks learn spatial-spectral fusion from data.
	These are mainly developed for remote sensing as this is a largely spread technological configuration.
	
	\paragraph{Ground-level unaligned RGB-guided HSI fusion.}
	Aligned RGB-guided HSR assumes that RGB and HSI pixels correspond to the same scene points, which is difficult to satisfy in real ground-level dual-camera systems. Fewer methods address this configuration compared to that of remote sensing.
	HSIFN~\cite{lai2025hsifn} explicitly considers real unaligned RGB guidance by aligning reference features before fusion.
	SSC-HSR~\cite{zhang2025sschsr} further combines two-stage alignment and feature aggregation with modulation, similarity, and attention-based fusion.
	These methods demonstrate the importance of registration-aware fusion, but rely on learned black-box pipelines that can be computationally demanding, less interpretable, and tied to training scale factors and spectral supports.
	Our work instead targets lightweight, interpretable model-based fusion robust to local registration failure and flexible across sensor supports.
	
	\paragraph{Cross-modal dense alignment.}
	Dense optical flow provides pixel-wise correspondences suitable for close-range RGB-HSI systems, where nonrigid misalignments such as parallax and depth discontinuities cannot be captured by a global affine transform.
	CMF~\cite{zhou2022promoting} extends the RAFT~\cite{teed2020raft} dense flow estimator to heterogeneous image pairs through modality-aware processing, making it suitable for images with different sensing characteristics.
	In this work, we use CMF to estimate correspondences between the HSI-derived RGB proxy and the RGB guide, which exhibit radiometric differences and cross-resolution bandwidth mismatch.
	Residual flow errors can nevertheless make these correspondences locally unreliable, motivating their explicit reliability assessment before RGB detail is injected.
	
	\paragraph{Registration-aware fusion and confidence.}
	Misalignment-aware guidance also appears in other multimodal problems, including RGB-guided thermal SR with edge-aware or unaligned guidance~\cite{gupta2020pyramidal,gupta2021toward}, RGB-guided IR resolution enhancement~\cite{trammer2023rgb}, and recent thermal SR benchmarks~\cite{rivadeneira2024thermal}.
	SuperFusion~\cite{tang2022superfusion}, for example, jointly considers registration and fusion using bidirectional deformation fields and symmetric constraints.
	First, we target HSR from real RGB-HSI VIS-NIR pairs.
	Second, rather than learning an end-to-end reconstruction, we use registration confidence explicitly as a pixel-wise weight map in spectral regression and in a gated detail injection fusion.
	
	\paragraph{Real RGB-HSI acquisition gap.}
	The closest standard evaluation protocol for unaligned RGB-guided HSR is the ``Real'' dataset~\cite{lai2025hsifn}.
	It uses HSI-HSI pairs from identical scan-based hyperspectral cameras; one HSI is spectrally projected to form the RGB guide and the other is spatially degraded, with evaluation restricted to 31 VIS bands. The pairs are manually cropped to their common support.
	This provides controlled quantitative evaluation but differs from a native RGB-HSI dual-camera acquisition.
	We therefore complement it with a real native RGB-HSI acquisition, acquired by our dual-camera setup, which is composed of a Raspberry Pi HQ RGB camera and the Ultris SR5 compact snapshot HSI camera.
	This setup exposes practical sensor effects absent from the Real protocol. We provide a 34-frame reduced-resolution quantitative evaluation and a native-resolution qualitative example over the full 51-band visible-near-infrared (VIS-NIR) support.

	\section{Problem Formulation and Notation}
	
	We aim to reconstruct a latent HR HSI
	\(\mathbf{X}\in\mathcal{X}\subseteq\mathbb{R}^{H\times W\times C}\),
	defined on the HSI grid (where $H$, $W$, and $C$ are the height, width, and number of spectral bands), from an observed LR HSI
	\(\mathbf{Y}\in\mathcal{Y}\subseteq\mathbb{R}^{h\times w\times C}\) (where $H > h$ and $W > w$)
	and an observed HR RGB image
	\(\mathbf{R}\in\mathcal{R}\subseteq\mathbb{R}^{H_R\times W_R\times3}\).
	We denote the reconstruction by \(\hat{\mathbf{X}}\), and the scale factor by
	\(\sigma=H/h=W/w\).
	
	The LR HSI observation is modeled as a spatially degraded version of the latent HR HSI:
	\begin{equation}
		\mathbf{Y}\approx\mathbf{D}_{\sigma}\mathcal{L}_{\sigma}\mathbf{X},%
		\label{eq:spatial_degradation}%
	\end{equation}%
	where \(\mathcal{L}_{\sigma}:\mathcal{X}\rightarrow\mathcal{X}\) is a spatial blur operator, representing the difference in spatial resolution between LR and HR, and
	\(\mathbf{D}_{\sigma}:\mathcal{X}\rightarrow\mathcal{Y}\) is a downsampling operator.

	The RGB observation is related to the latent HSI through spectral projection and geometric correspondence. Let
	\(\phi_{H\rightarrow R}: \Omega_H \subset \mathbb{R}^{2} \rightarrow \Omega_R \subset \mathbb{R}^{2}\) denote the unknown coordinate map from the HSI grid to the RGB grid. Ideally, for each HSI-grid location \(x\in\Omega_H\),
	\begin{equation}
		\mathbf{R}\!\left(\phi_{H\rightarrow R}(x)\right)
		\approx
		\Psi(\mathbf{X})(x),%
		\label{eq:spectral_degradation}%
	\end{equation}%
	where \(\Psi:\mathcal{X}\rightarrow\mathbb{R}^{H\times W\times3}\) maps an HSI cube to an RGB proxy.
	In practice, $\Psi$ is the fixed HSI-to-RGB spectral projection used by the corresponding experimental pipeline.
	
	In practice, \(\phi_{H\rightarrow R}\) is unknown and is parametrized by a dense flow field
	\(\mathbf{F}_{H\rightarrow R}\in\mathbb{R}^{H\times W\times2}\) between the HSI-derived RGB proxy
	\(\Psi(\mathbf{H}^{\uparrow})\) and the observed RGB image \(\mathbf{R}\), where
	\(\mathbf{H}^{\uparrow}\in\mathbb{R}^{H\times W\times C}\) denotes the LR HSI upsampled to the target HR HSI grid, for example by bicubic interpolation, but without the HR spatial details.
	The flow parametrizes the coordinate map, in the chosen sampling coordinates, as
	$\phi_{H\rightarrow R}(x) = x+\mathbf{F}_{H\rightarrow R}(x)$, with $x\in\Omega_H$.
	Accordingly, the RGB guide warped onto the HSI grid is
	\(\mathbf{R}_w(x) = \mathbf{R}\!\left(\phi_{H\rightarrow R}(x)\right) = \mathbf{R}(x+\mathbf{F}_{H\rightarrow R}(x))\).
	
	Given $\mathbf{Y}$ and the registered RGB guide $\mathbf{R}_w$, the RGB-guided HSR problem is to reconstruct:
	\begin{equation}
	\hat{\mathbf{X}}
	=
	\mathcal{S}_{\sigma}\!\left(\mathbf{Y},\mathbf{R}_w\right),
	\end{equation}
	Ideally, it should satisfy the HSI degradation model \(\mathbf{Y}\approx\mathbf{D}_{\sigma}\mathcal{L}_{\sigma}\hat{\mathbf{X}}\) and, where RGB correspondence is reliable, the registered RGB observation \(\Psi(\hat{\mathbf{X}})(x)\approx\mathbf{R}_w(x)\), with $x\in\Omega_H$.

	The remaining issue is that \(\mathbf{R}_w\) is not uniformly reliable. Even when the sampling coordinate lies inside the RGB image, the correspondence may be incorrect near occlusions, depth discontinuities, non-overlapping regions, or weak cross-modal texture. The proposed method therefore estimates both a warped RGB guide and a lightweight confidence map
	\(\mathbf{C}\in[0,1]^{H\times W}\), used as a reliability map by the reconstruction operator
	\(\mathcal{S}_{\sigma}\).

\section{Methodology}
	
\figurename~\ref{fig:pipeline} summarizes our instantiation of the reconstruction operator \(\mathcal{S}_{\sigma}\). The method first estimates a dense HSI-to-RGB correspondence and back-warps the RGB image to the HSI grid. It then estimates the reliability of the warped RGB guide, uses this confidence inside the GSA spectral regression, and finally gates the RGB-guided detail residual.

	\subsection{Cross-modal flow alignment}
	
	We estimate the coordinate map \(\phi_{H\rightarrow R}\) through its flow representation, using a dense flow estimator \(\mathcal{F}\) applied to the HSI-derived RGB proxy and the observed RGB guide:
	\begin{equation}
		\mathbf{F}_{H\rightarrow R}
		=
		\mathcal{F}
		(
		\Psi(\mathbf{H}^{\uparrow}),\mathbf{R}
		),
		\qquad
		\mathbf{F}_{H\rightarrow R}\in\mathbb{R}^{H\times W\times2},
		\label{eq:flow}
	\end{equation}
	where \(\mathcal{F}\) is practically instantiated by CMF/CrossRAFT~\cite{zhou2022promoting}, a recent cross-modal optical-flow estimator designed to promote single-modal optical-flow networks to heterogeneous image pairs. Other dense registration modules could be used in the same formulation.
	
	The RGB image is then resampled on the HSI grid by backward warping:
	\begin{equation}
		\mathbf{R}_w
		=
		\mathcal{B}_{\mathbf{F}_{H\rightarrow R}}
		(
		\mathbf{R}
		)
		\quad
		\Longleftrightarrow
		\quad
		\mathbf{R}_w(x)
		=
		\mathbf{R}\!\left(x+\mathbf{F}_{H\rightarrow R}(x)\right),
		\quad
		x\in\Omega_H,
		\label{eq:warp}
	\end{equation}
	where \(\mathcal{B}_{\mathbf{F}_{H\rightarrow R}}\) samples \(\mathbf{R}\) at the RGB-grid coordinates \(x+\mathbf{F}_{H\rightarrow R}(x)\) for each \(x\in\Omega_H\), using bilinear interpolation in practice.
	A direct registration-and-fusion baseline is therefore
	\begin{equation}
		\hat{\mathbf{X}}_{\mathrm{CMF\text{-}GSA}}
		=
		\operatorname{GSA}_{\sigma}
		(
		\mathbf{H}^{\uparrow},
		\mathbf{R}_w
		).
		\label{eq:baseline_cmf_gsa}
	\end{equation}
	This baseline assumes that every valid warped RGB pixel is equally trustworthy. The proposed method replaces this uniform-trust assumption with a spatial confidence map.
	Dense flow handles spatial parallax, while true occlusions
	and non-overlapping fields of view
	cannot be registered and are instead handled by the validity mask and reliability model below.
	
	\subsection{Confidence map and reliability cues}
	
	We estimate a confidence map \(\mathbf{C}\in[0,1]^{H\times W}\) that measures whether the warped RGB guide should be trusted locally during fusion. The principle is to assign high confidence only to pixels that are geometrically valid and locally consistent with the HSI-derived RGB proxy. As shown in the following, the confidence is not a learned uncertainty model; it is an energy-based reliability score used as a spatial weight.
	
	\paragraph{Validity mask.}
	A binary geometric validity mask \(\mathbf{M}\in\{0,1\}^{H\times W}\) is obtained from the warping grid. For each \(x\in\Omega_H\), we set \(\mathbf{M}(x)=1\) if the sampled coordinate \(x+\mathbf{F}_{H\rightarrow R}(x)\) lies inside the RGB image domain \(\Omega_R\), and \(\mathbf{M}(x)=0\) otherwise. This mask removes pixels outside the intersection of the fields of view of the two cameras.
	Equivalently, the validity mask $\mathbf{M}$ is obtained as follows:
	\begin{equation}
	\mathbf{M}(x)
	=
	\begin{cases}
		1, & x+\mathbf{F}_{H\rightarrow R}(x)\in\Omega_R,\\
		0, & \text{otherwise}.
	\end{cases}
	\label{eq:valid_mask}
	\end{equation}
	
	\paragraph{RGB proxy residual.}
	The first cue compares the HSI-derived RGB proxy with a low-pass version of the warped RGB guide, so that both terms have comparable spatial bandwidth:
	\begin{equation}
		E_{\mathrm{rgb}}(x)
		=
		\frac{1}{3}
		\sum_{c=1}^{3}
		\left|
		\Psi(\mathbf{H}^{\uparrow})_c(x)
		-
		\mathcal{L}_{R}(\mathbf{R}_w)_c(x)
		\right|.
		\label{eq:rgb_residual}
	\end{equation}
	Here, \(\mathcal{L}_{R}\) denotes the RGB low-pass operator used to match the bandwidth of the HSI proxy. We use an \(\ell_1\) residual because it is less sensitive than a squared residual to local radiometric mismatch and outliers.
	This cue is appropriate when the proxy and guide are radiometrically
	comparable, as on Real where both are derived from HSI-domain data.

In real snapshot dual-camera systems, independent sensor responses, exposure, and high HSI noise can keep $E_{\rm rgb}$ high even under correct alignment, as for Ultris-RPi. We therefore use a structure-consistency residual based on local zero-normalized cross-correlation (ZNCC)~\cite{di2005zncc}.
We denote two scalar images
\(
U=\mathcal{G}_{s}\!\left(\ell(\Psi(\mathbf{H}^{\uparrow}))\right),
\)
and
\(
V=\mathcal{G}_{s}\!\left(\ell(\mathbf{R}_w)\right),
\)
where \(\ell(\cdot)\) is a fixed RGB-to-luminance conversion and \(\mathcal{G}_{s}\) is a Gaussian
smoothing operator. For two scalar images \(U\) and \(V\), the local ZNCC over a window \(\Omega_x\) is:
{\footnotesize
\begin{equation}
	\operatorname{ZNCC}_{\Omega_x}(U,V)
	=
	\frac{
		\sum_{u\in\Omega_x}
		\left(U(u)-\mu_{U, x}\right)
		\left(V(u)-\mu_{V, x}\right)
	}{
		\sqrt{
			\sum_{u\in\Omega_x}
			\left(U(u)-\mu_{U, x}\right)^2
		}
		\sqrt{
			\sum_{u\in\Omega_x}
			\left(V(u)-\mu_{V, x}\right)^2
		}
		+\epsilon
	},
	\label{eq:zncc}
\end{equation}
}
where \(\mu_{U, x}\) and \(\mu_{V, x}\) are local means over \(\Omega_x\), and \(\epsilon>0\) avoids
numerical instability. The structure residual is then:
\begin{equation}
	E_{\mathrm{str}}(x)
	=
	\mathbf{M}_{\mathrm{str}}(x)
	\left[
	1-
	\max
	\left(
	0,
	\operatorname{ZNCC}_{\Omega_x}(U,V)
	\right)
	\right],
	\label{eq:zncc_residual}
\end{equation}
where \(\mathbf{M}_{\mathrm{str}}(x)=1\) only when both local standard deviations are above a
minimum threshold, and \(\mathbf{M}_{\mathrm{str}}(x)=0\) otherwise. Thus, flat or low-texture regions
are treated as uninformative rather than unreliable.
	
\paragraph{Cycle consistency.}
If bidirectional flows are available, a cycle-consistency error can complement the above cues:
\begin{equation}
	E_{\mathrm{cyc}}(x)
	=
	\left\|
	\mathbf{F}_{H\rightarrow R}(x)
	+
	\mathcal{B}_{\mathbf{F}_{H\rightarrow R}}
	\left(
	\mathbf{F}_{R\rightarrow H}
	\right)(x)
	\right\|_2 .
	\label{eq:cycle_residual}
\end{equation}
This cue penalizes forward-backward disagreement and is expected to identify occlusions and inconsistent correspondences.

\begin{figure}[t]
	\centering
	\includegraphics[width=0.5\linewidth]{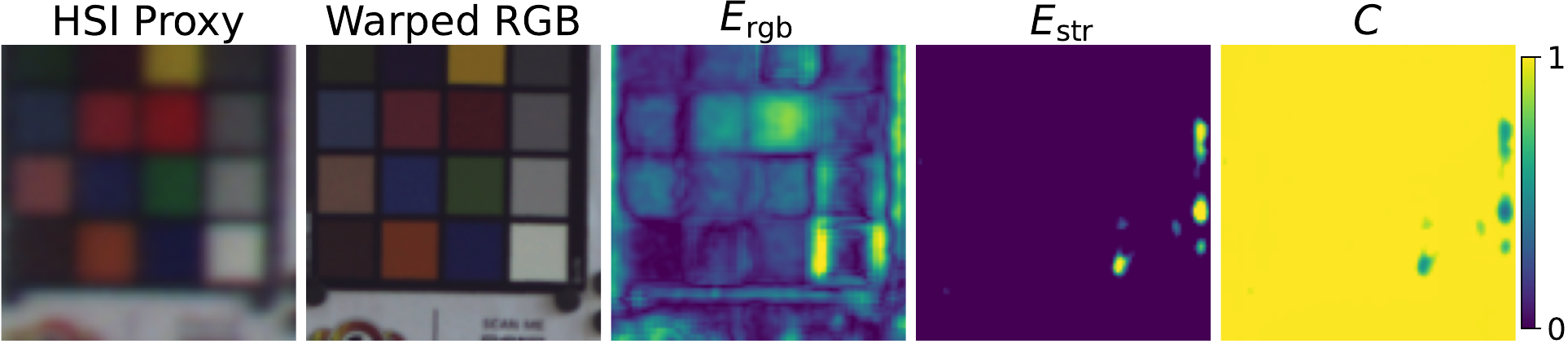}%
	\caption{Confidence cues on a native Ultris-RPi crop. From left to right:
		HSI proxy, warped RGB, $E_{\rm rgb}$, $E_{\rm str}$, and confidence $C$.
		The residual maps are percentile-normalized for visualization, while
		$C$ is shown on its native $[0,1]$ scale.}%
	\label{fig:confidence_cues}%
\end{figure}

\paragraph{Confidence map.}
We define the local reliability energy as:
\begin{equation}
	\mathcal{E}(x)
	=
	\alpha E_{\mathrm{rgb}}(x)
	+
	\beta E_{\mathrm{str}}(x)
	+
	\gamma E_{\mathrm{cyc}}(x),
	\label{eq:confidence_energy}
\end{equation}
where \(\alpha,\beta,\gamma\ge0\) control the rejection strengths of raw RGB-proxy inconsistency, local structure inconsistency, and cycle-flow disagreement, respectively. These parameters are selected on the validation set when a reference is available, or fixed for real-device evaluation. The confidence is:
\begin{equation}
	\mathbf{C}(x)
	=
	\mathbf{M}(x)\exp[-\mathcal{E}(x)].
	\label{eq:confidence}
\end{equation}
Although the gate is pixel-wise, $E_{\rm rgb}$ uses bandwidth-matched images and $E_{\rm str}$ is computed over a smoothed local window, so confidence is spatially supported; sharp drops are expected mainly near local misregistration or parallax, where suppressing RGB detail prevents inconsistent structures from being transferred.
A native Ultris-RPi example of these cues and the resulting confidence is
shown in \figurename~\ref{fig:confidence_cues}.
The exponential maps additive reliability penalties to a multiplicative weight in \([0,1]\); each cue can only decrease the trust assigned to the warped RGB guide. This confidence is not a calibrated posterior probability; it is used as a spatial precision weight for regression and as a gate for detail injection.

\subsection{Confidence-weighted Gram-Schmidt adaptive fusion}

We propose a method based on GSA, which is a well known technique for the fusion of LR multispectral (e.g., RGB) and HR panchromatic (i.e., single band) satellite remote sensing images~\cite{aiazzi2007improving}. 
Specifically, we extend the GSA method to address RGB-HSI fusion.
The proposed GSA method estimates the relation between the HSI spectral bands and the low-pass RGB guide, then injects the high-frequency component of the RGB guide into the HSI reconstruction. When the RGB image is replaced by a panchromatic one, this corresponds to the standard 
GSA algorithm~\cite{aiazzi2007improving}.

Conversely to standard GSA, here we consider a penalized multivariate regression between the three RGB channels and the HSI bands.
Let \(\mathbf{A}\in\mathbb{R}^{N\times C}\) denote centered HSI samples extracted from \(\mathbf{H}^{\uparrow}\), and let \(\mathbf{Q}\in\mathbb{R}^{N\times3}\) denote centered samples from the low-pass component of the warped RGB guide. Let \(w_i=\mathbf{C}(x_i)\) be the confidence value at sample location \(x_i\). We estimate the regression matrix \(\boldsymbol{\Theta}\in\mathbb{R}^{C\times3}\) by solving
\begin{equation}
	\boldsymbol{\Theta}^{\star}
	=
	\arg\min_{\boldsymbol{\Theta}}
	\sum_{i=1}^{N}
	w_i
	\left\|
	\mathbf{Q}_i
	-
	\mathbf{A}_i\boldsymbol{\Theta}
	\right\|_2^2
	+
	\lambda
	\left\|
	\boldsymbol{\Theta}
	\right\|_F^2 .
	\label{eq:wls_regression}
\end{equation}
In matrix form, with
\(\boldsymbol{\Lambda}=\operatorname{diag}(w_1,\ldots,w_N)\), this gives
\begin{equation}
	\boldsymbol{\Theta}^{\star}
	=
	\left(
	\mathbf{A}^{\top}
	\boldsymbol{\Lambda}
	\mathbf{A}
	+
	\lambda\mathbf{I}
	\right)^{-1}
	\mathbf{A}^{\top}
	\boldsymbol{\Lambda}
	\mathbf{Q}.
	\label{eq:wls_solution}
\end{equation}
The Tikhonov parameter \(\lambda\) stabilizes the regression, especially in the presence of high-frequency noise, which is typical in snapshot HSI acquisitions.
The regression predicts the HSI-explained low-frequency RGB component $\mathbf{I}=\mathbf{A}\boldsymbol{\Theta}^{\star}$.
The residual RGB detail is then $\mathbf{D}=\mathbf{R}_w-\mathbf{I} \in \mathbb{R}^{N\times3}$, and is injected through band-wise GSA gains $\mathbf{S}\in\mathbb{R}^{3\times C}$ estimated from $\mathbf{A}$ and $\mathbf{I}$, yielding:
\begin{equation}
	\widehat{\mathbf{X}}_G
	=
	\mathbf{H}^{\uparrow}
	+
	\mathbf{D}\mathbf{S}.
\end{equation}

\begin{figure}[t]
	\centering
	\includegraphics[width=0.26\linewidth]{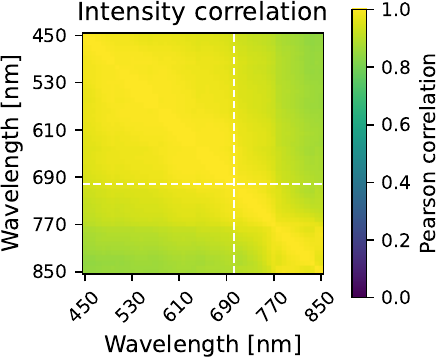}%
	\hfill
	\includegraphics[width=0.26\linewidth]{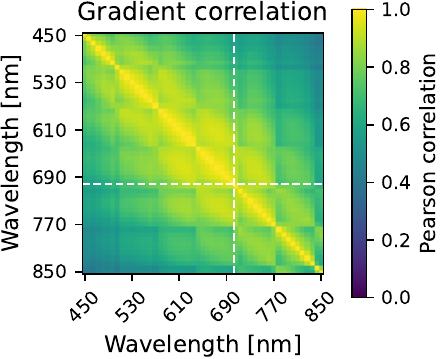}%
	\hfill
	\includegraphics[width=0.20\linewidth]{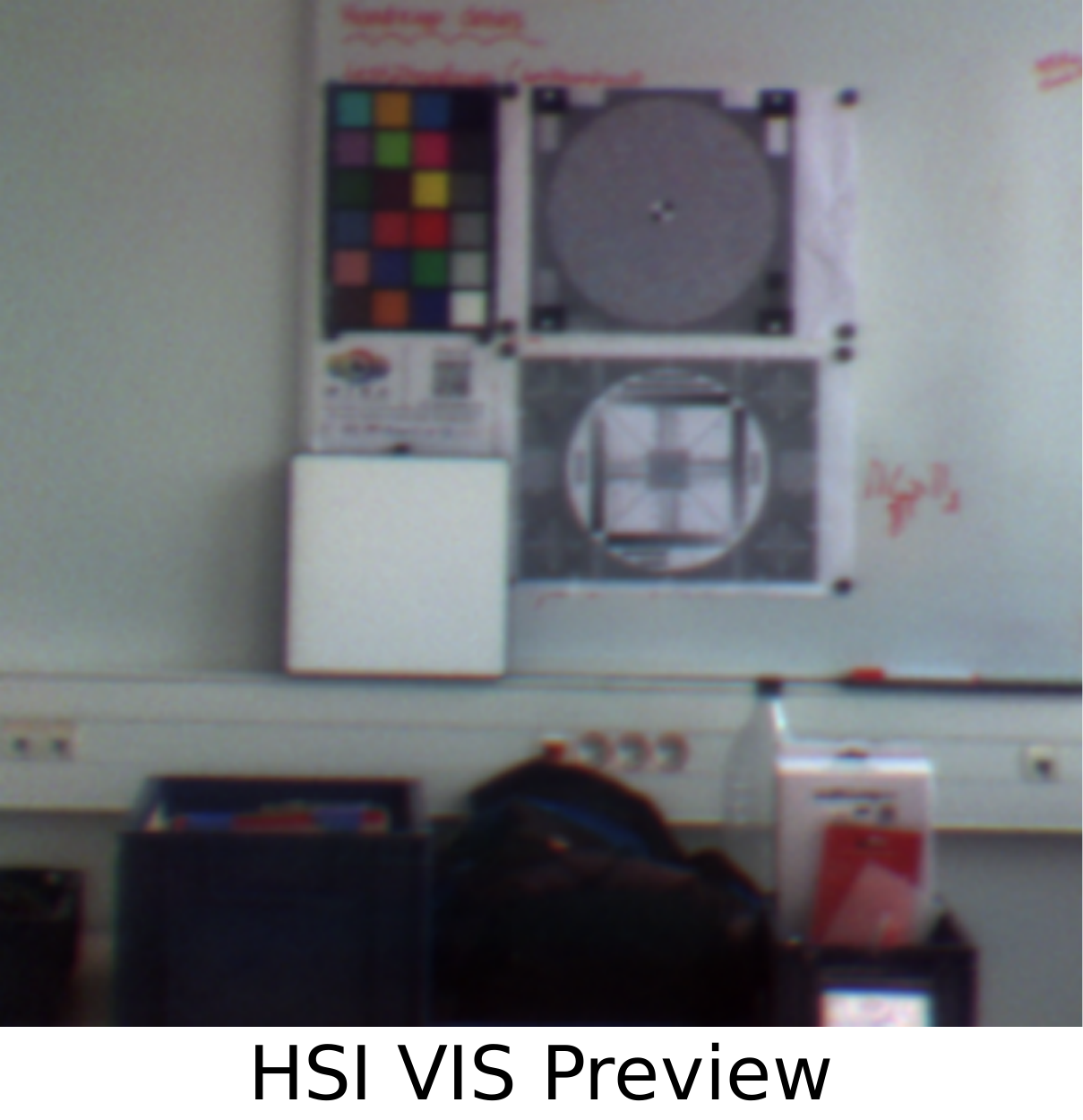}%
	\hfill
	\includegraphics[width=0.20\linewidth]{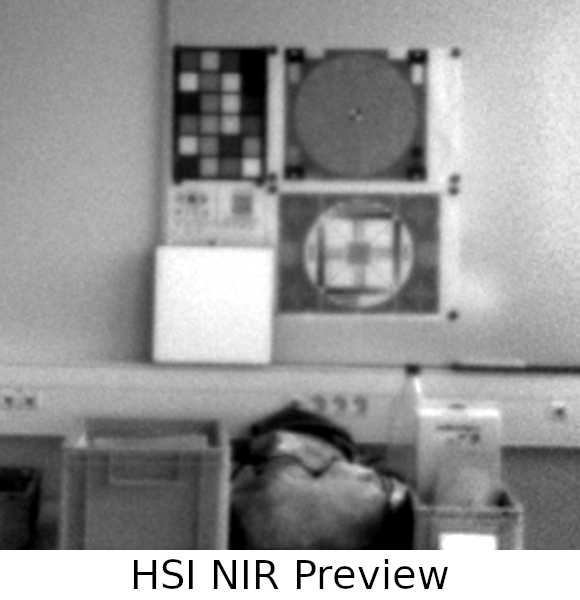}%
	\caption{VIS-NIR band correlation on an Ultris HSI. From left to right: intensity correlation, gradient correlation, and VIS and NIR previews of the same HSI. Dashed lines mark the 700\,nm VIS-NIR boundary.}%
	\label{fig:vis_nir_correlation}%
\end{figure}

Thus, RGB guidance supplies spatial detail rather than predicting NIR
radiance, as the bands remain anchored to $\mathbf{H}^{\uparrow}$,
while the RGB residual is transferred through the HSI-derived gains
$\mathbf{S}$. This transfer is meaningful when spatial structures remain
correlated across the VIS-NIR boundary, as observed in our Ultris data in
\figurename~\ref{fig:vis_nir_correlation}.

This estimate contains spatial details from the RGB guide, but those details may still be unreliable in low-confidence regions. Therefore, we apply a final gating step.

\subsection{Confidence-gated detail injection}

The confidence map is finally used to control how much RGB-induced detail is retained in the reconstruction. We combine two candidate estimates with complementary behavior. The first one is the upsampled HSI \(\mathbf{H}^{\uparrow}\), which preserves the spectral content of the observed HSI but remains spatially smooth. The second one is the confidence-weighted GSA estimate \(\hat{\mathbf{X}}_{G}\), which contains high-frequency detail transferred from the registered RGB guide but may still be locally unreliable when the RGB-HSI correspondence is incorrect.

We use the confidence map as a spatially varying convex gate between these two estimates:
\begin{equation}
	\hat{\mathbf{X}}(x)
	=
	(1-\mathbf{C}(x))\mathbf{H}^{\uparrow}(x)
	+
	\mathbf{C}(x)\hat{\mathbf{X}}_{G}(x),
	\qquad
	\mathbf{C}(x)\in[0,1].
	\label{eq:final_gate}
\end{equation}
The confidence is broadcast along the spectral dimension. Thus, when \(\mathbf{C}(x)=0\), the reconstruction falls back to the HSI-preserving estimate; when \(\mathbf{C}(x)=1\), it uses the RGB-guided GSA estimate; intermediate values attenuate the RGB-induced detail.

Equivalently, Eq.~\eqref{eq:final_gate} can be written as gated detail injection:
\begin{equation}
	\hat{\mathbf{X}}
	=
	\mathbf{H}^{\uparrow}
	+
	\mathbf{C}
	\odot
	\left(
	\hat{\mathbf{X}}_{G}
	-
	\mathbf{H}^{\uparrow}
	\right).
	\label{eq:gated_detail}
\end{equation}
This form makes explicit that the confidence gate does not replace the hyperspectral estimate. It only controls the amount of RGB-guided high-frequency detail added to it.

The gate also admits a local least-squares interpretation. For each spatial location \(x\), Eq.~\eqref{eq:final_gate} is the minimizer of:
{\small
\begin{equation}
	E_x(\mathbf{Z})
	=
	\frac{1-\mathbf{C}(x)}{2}
	\left\|
	\mathbf{Z}
	-
	\mathbf{H}^{\uparrow}(x)
	\right\|_2^2
	+
	\frac{\mathbf{C}(x)}{2}
	\left\|
	\mathbf{Z}
	-
	\hat{\mathbf{X}}_{G}(x)
	\right\|_2^2 .
	\label{eq:local_ls_gate}
\end{equation}
}
Therefore, the final reconstruction is a reliability-weighted least-squares compromise between a spectrally conservative HSI estimate and an RGB-guided detail-enhanced estimate.

	\section{Experiments}
	\label{sec:experiments}
	
	\subsection{Experimental setup}
	
	\subsubsection{Datasets}
	
	\paragraph{Real dataset.} We first evaluate on the Real unaligned RGB-guided HSR benchmark~\cite{lai2025hsifn}. The dataset contains 57 paired hyperspectral acquisitions with real non-rigid misalignment.
	One HSI is spectrally projected to simulate the HR RGB guide,
	and the other is spatially degraded according to Eq.~\eqref{eq:spatial_degradation}, using a Gaussian blur with standard deviation $\sigma/2$ followed by area-based downsampling, with the original HSI used as the HR reference.
	All methods are re-evaluated using these common inputs and a valid-mask evaluation described in Sec.~\ref{sec:metrics}. Consequently, baseline values may differ slightly from those originally reported.
	We use the 31-band visible setting and evaluate scale factors $\times4$ and $\times8$ on 10 test frames. Hyperparameters are selected on the 47-frame validation split and then fixed for test evaluation.

\begin{figure}[t]
		\centering
		\begin{minipage}[b]{0.29\linewidth}
			\centering
			\includegraphics[width=\linewidth]{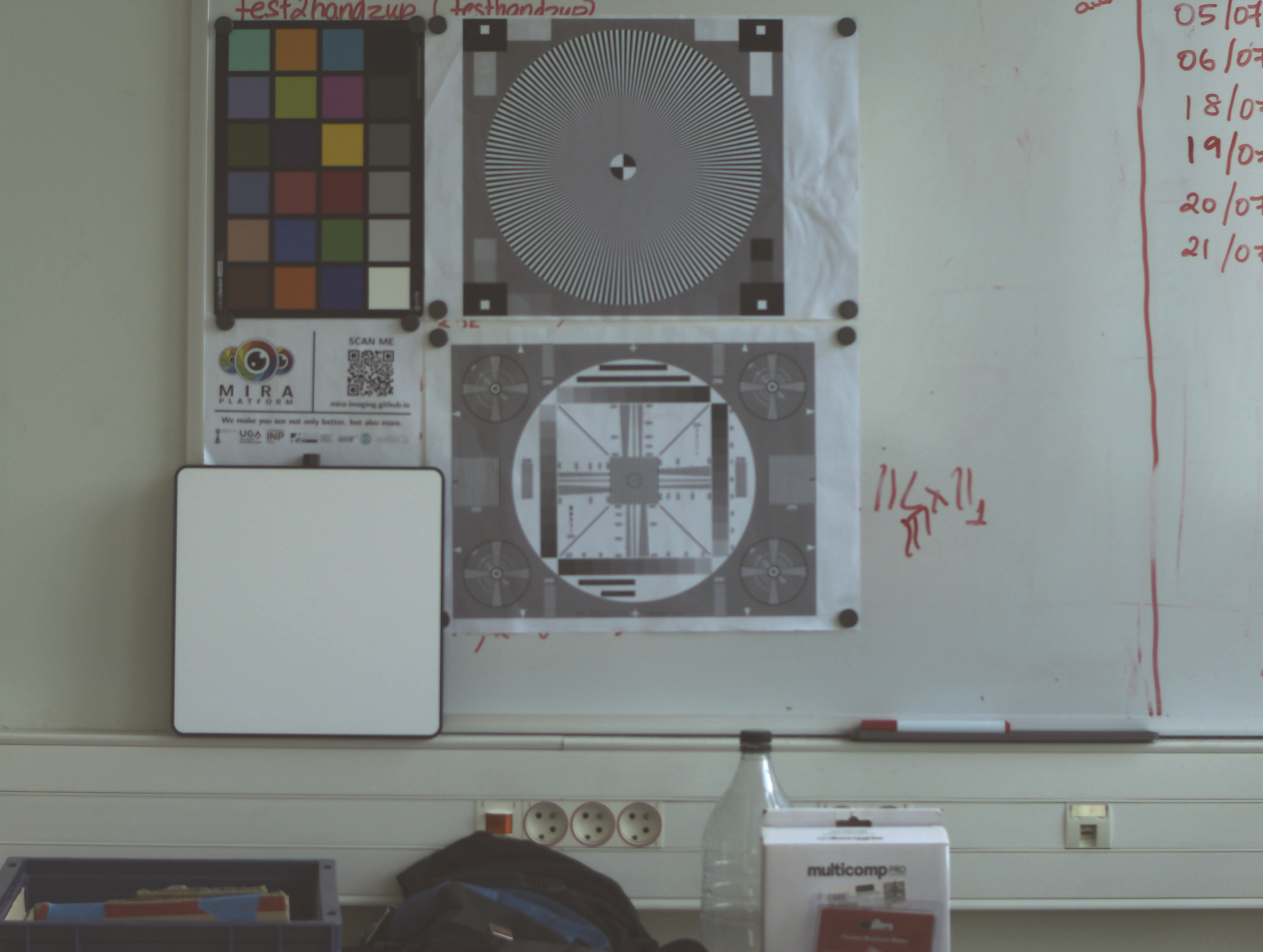}\\
			{\small \textbf{(a)} RGB camera view}
		\end{minipage}%
		\hfill
		\begin{minipage}[b]{0.47\linewidth}
			\centering
			\includegraphics[width=0.9\linewidth]{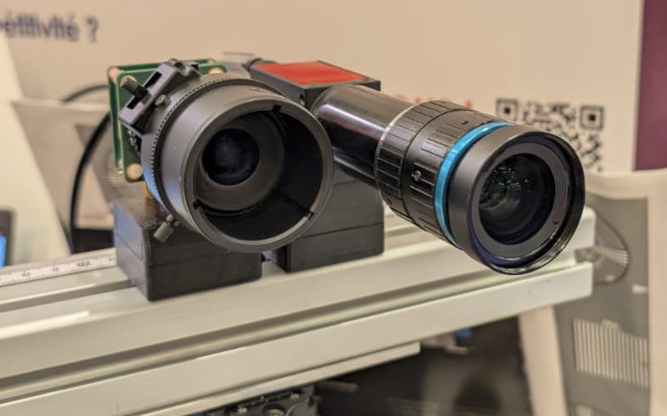}\\
			{\small \textbf{(b)} Acquisition setup}
		\end{minipage}%
		\hfill
		\begin{minipage}[b]{0.23\linewidth}
			\centering
			\includegraphics[width=\linewidth]{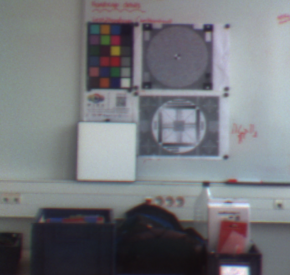}\\
			{\footnotesize \textbf{(c)} HSI camera view}
		\end{minipage}%
		\caption{%
			Ultris-RPi acquisition system used for the real-device evaluation, composed of a Raspberry Pi HQ RGB camera (left) and an Ultris SR5 compact snapshot hyperspectral camera (right), mounted on a rail. The corresponding observed RGB and HSI images are shown after flat-field correction with the white reference. The two cameras have different fields of view, spatial resolutions, and sensor and noise characteristics.%
		}%
		\label{fig:ultris_rpi_setup}%
	\end{figure}

	\paragraph{Ultris-RPi real-device acquisition.}
	To assess deployment beyond the Real benchmark, we additionally evaluate an Ultris-RPi sequence acquired with the Ultris SR5 compact snapshot hyperspectral camera and a Raspberry Pi HQ RGB camera, shown in \figurename~\ref{fig:ultris_rpi_setup}.
	Unlike Real, whose RGB guidance is synthesized from HSI-HSI pairs, this is a true dual-camera RGB-HSI acquisition with different fields of view, sensor responses, spatial resolutions, and noise characteristics.
	Since simultaneous HR HSI ground truth is unavailable in this real setup, we use a reduced-resolution quantitative evaluation \cite{ciotola2024hyperspectral} on a 34-frame static sequence, using the observed HSI as reference, and synthetically degrading it by a factor $\times4$.
	We additionally report a qualitative analysis on a single frame in its native resolution.
	SSC-HSR is evaluated on the 31-band VIS subset, while CMF-GSA-wg is evaluated on both the same 31-band subset and the full 51-band VIS-NIR cube.
	
	\subsubsection{Compared methods}
	
	We compare against a learning-based SISR baseline, ESSAformer~\cite{zhang2023essaformer}, and two learning-based unaligned RGB-guided HSR baselines, HSIFN~\cite{lai2025hsifn} and SSC-HSR~\cite{zhang2025sschsr}. We also compare several variants of our lightweight CMF-GSA hybrid pipeline. CMF-GSA denotes CMF alignment followed by valid-mask GSA. CMF-GSA-w uses the confidence map only as weighted least-squares weights in the GSA spectral regression. CMF-GSA-wg is the full method, using confidence both in the regression and in the final two-expert detail gate.

	On Ultris-RPi, a reduced-resolution quantitative evaluation and a native-resolution qualitative evaluation are provided.
	SSC-HSR is evaluated in its 31-band VIS subset,
	whereas CMF-GSA-wg is evaluated both on the same 31-band subset, providing a direct comparison, and on the full 51-band VIS-NIR cube.
	We do not evaluate 31-band learned baselines in the 51-band setting without retraining or spectral-domain adaptation~\cite{picone2026leveraging}.

\subsubsection{Metrics}
\label{sec:metrics}
Let \(\mathbf{X}^{\star},\hat{\mathbf{X}}\in\mathbb{R}^{H\times W\times C}\) denote the reference and reconstructed HSIs. Metrics are computed on the valid spatial support \(\Omega\) induced by the registration mask, and all images are evaluated in the data range \([0,1]\). We report RMSE, mean PSNR (mPSNR), mean SSIM (mSSIM), and spectral angle mapper (SAM). RMSE is computed globally over all valid pixels and spectral bands, while mPSNR and mSSIM are averaged over bands. SAM is reported as the average spectral angle over valid pixels. In addition to scalar metrics, we visualize pixel-wise RMSE and SAM error maps for selected frames.
All methods are evaluated on the HSI grid. Quantitative metrics are computed on the same valid spatial support for a given frame, induced by the CMF mask, so that no method is favored by a different border region. We report runtime on a Linux workstation with an NVIDIA RTX 3500 Ada Generation Laptop GPU, an x86-64 CPU, and PyTorch 2.11.0 with CUDA 13.0.

\subsubsection{Hyperparameter selection}

For the Real dataset, hyperparameters are selected on the full 47-frame validation split and
then fixed for test evaluation. We perform a grid search over the confidence parameters and
the GSA Tikhonov weight, using a joint RMSE-SAM validation criterion. The selected values
are
$
(\alpha,\beta,\gamma,\lambda)=(25,0,0.1,10^{-4})
$
for \(\times4\), and
$
(\alpha,\beta,\gamma,\lambda)=(15,0,0.1,10^{-4})
$
for \(\times8\). Thus, on Real, the raw RGB proxy residual is the main confidence cue,
while cycle consistency acts as a moderate auxiliary cue and the ZNCC structure cue is disabled.

For Ultris-RPi, no reference-based hyperparameter search is performed.
The raw RGB proxy residual \(E_{\mathrm{rgb}}\) is disabled because, in a real dual-camera acquisition, it responds not only to misregistration but also to cross-sensor radiometry, HSI noise, and imperfect spatial-bandwidth matching.
For Ultris-RPi, the structural reliability term is instantiated as a local
ZNCC cue computed after Gaussian smoothing.
For both Ultris-RPi settings we use $(\alpha,\beta,\gamma,\lambda)=(0,10,0.01,0.1)$. The ZNCC structural cue uses a $5\times5$ window, minimum local standard deviation $0.01$, and threshold $0.05$;
the Gaussian smoothing standard deviation is $0.7$ for reduced-resolution
evaluation and $1.0$ for the native-resolution evaluation.

	\subsection{Real benchmark evaluation}

	\paragraph{Quantitative comparison.} \tablename~\ref{tab:real_main_results} reports the main comparison on the Real test set.
	The full confidence-gated CMF-GSA-wg outperforms the learned RGB-guided baselines at both scale factors and the HSI-only ESSAformer SISR baseline at \(\times4\), while remaining substantially faster.
	At $\times4$, CMF-GSA-wg obtains the best RMSE, mPSNR, mSSIM, and SAM,
	with values of $0.0168$, $37.34$ dB, $0.94$, and $2.17^\circ$,
	respectively.
	At $\times8$, CMF-GSA-wg again achieves the best four reconstruction metrics, including RMSE $0.0257$ and SAM $2.94^\circ$.
	
	The comparison also shows that weighted regression alone does not explain the main gain. CMF-GSA-w improves SAM and mSSIM over valid-mask CMF-GSA, but gives only marginal changes in RMSE and mPSNR. The major improvement comes from the final two-expert gate, which suppresses unreliable RGB detail injection. At $\times4$, CMF-GSA-wg reduces RMSE by about $33\%$ relative to CMF-GSA-w.
	
	\tablename~\ref{tab:efficiency} further reports learned parameter counts and peak GPU memory.
	All $39.76$M parameters of CMF-GSA-wg belong to the frozen CMF registration backbone,
	while the confidence estimation and SR/fusion stages add no learned parameters.
	Despite the larger registration backbone, CMF-GSA-wg has the lowest peak memory at $0.69$ GiB.

	\paragraph{Confidence-cue ablation}
	
	\tablename~\ref{tab:confidence_ablation} isolates the confidence cues in the full gated model.
	Removing the RGB residual cue, i.e. setting $\alpha=0$, consistently degrades RMSE, mPSNR, mSSIM, and SAM at both scale factors, confirming that RGB-HSI proxy consistency is the dominant reliability cue on the Real benchmark.
	Removing cycle consistency, i.e. setting $\gamma=0$, gives metrics nearly tied to the full model, while reducing runtime because the backward flow no longer needs to be estimated.
	This supports our interpretation of cycle consistency as an auxiliary diagnostic cue in this cross-resolution setting.

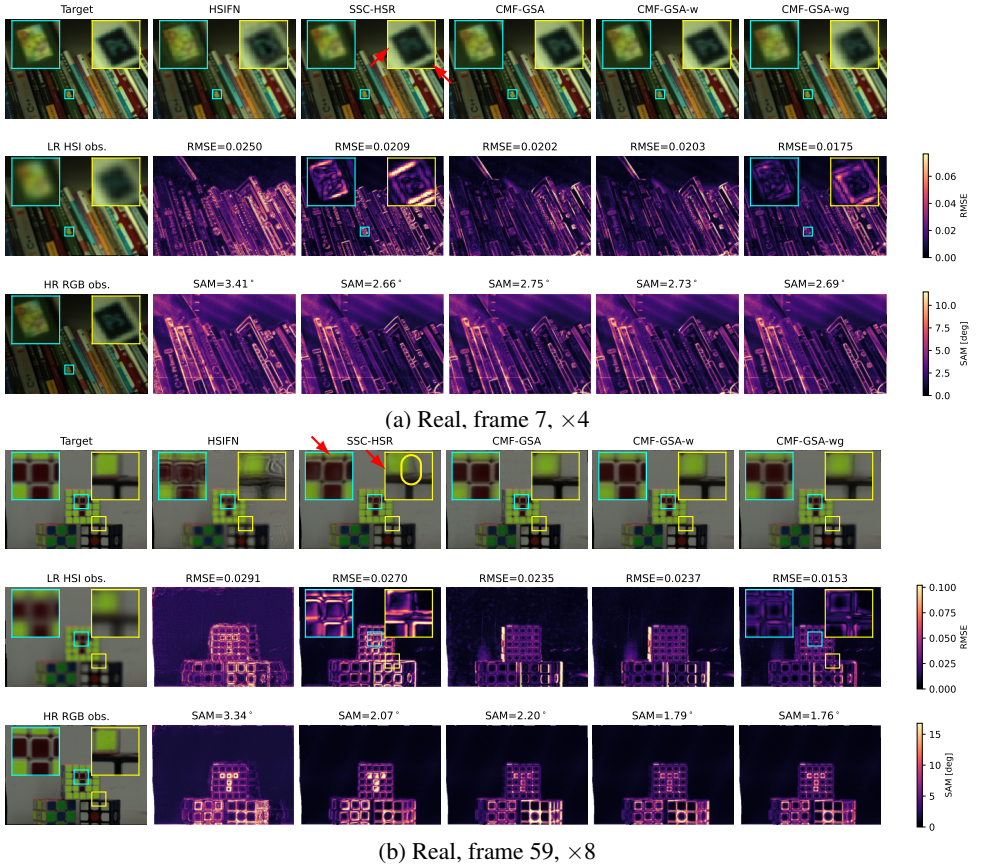
\begin{figure*}[!t]
	\centering
	\begin{minipage}{\textwidth}%
		\centering
		\begin{overpic}[width=\linewidth]{figures/experiments/real/sf4.00/qualitative/real\_frame\_7\_main\_qualitative.pdf}%
			\put(38,35){%
				\thicklines
				\color{red}
				\vector(1,1){2}
			}%
			\put(46.3,33.3){%
				\thicklines
				\color{red}
				\vector(-1,1){2}
			}%
		\end{overpic}%
	\end{minipage}\\
	{\small(a) Real, frame 7, $\times4$}%
	\\[0.2em]
	\begin{minipage}{\textwidth}%
		\centering
		\begin{overpic}[width=\linewidth]{figures/experiments/real/sf8.00/qualitative/real\_frame\_59\_main\_qualitative.pdf}%
			\put(42.2,37.9){%
				\rotatebox{0}{%
					\begin{picture}(0,0)
						\thicklines
						\color{yellow}
						\put(0,0){\oval(2,3)}
					\end{picture}%
				}%
			}%
			\put(37.6,39.9){%
				\thicklines
				\color{red}
				\vector(1,-1){2}
			}%
			\put(32.0,41.3){%
				\thicklines
				\color{red}
				\vector(1,-1){2}
			}%
		\end{overpic}%
	\end{minipage}\\
	{\small(b) Real, frame 59, $\times8$}%
	\caption{%
		Qualitative comparison on the Real test set at
		(a) \(\times4\) and (b) \(\times8\).
		The first column shows the target, LR HSI observation, and HR RGB observation;
		method columns show RGB previews, RMSE maps, and SAM maps.
		Two regions are highlighted for local inspection, with corresponding RMSE
		enlargements for SSC-HSR and CMF-GSA-wg.
		At \(\times4\), the visual differences are relatively subtle but CMF-GSA-wg
		reduces local RMSE in the highlighted regions.
		At \(\times8\), over-enhanced or spatially shifted structures are more apparent
		in SSC-HSR, whereas CMF-GSA-wg reduces the corresponding local error.
		In both examples, CMF-GSA-wg obtains the lowest RMSE and comparable or lower SAM.
		RMSE insets use a shared local scale for SSC-HSR and CMF-GSA-wg,
		separate from the full-map colorbar.%
	}%
	\label{fig:real_qualitative}%
\end{figure*}

\paragraph{Qualitative analysis.} \figurename~\ref{fig:real_qualitative} shows representative Real test examples at $\times4$ and $\times8$, respectively.
The selected examples support the quantitative results. At $\times4$, the visual differences are subtle but CMF-GSA-wg reduces local RMSE in the highlighted regions. At $\times8$, SSC-HSR shows more apparent over-enhanced or displaced structures, while the proposed gate suppresses unreliable RGB detail injection. In both frames, CMF-GSA-wg obtains the lowest RMSE and comparable or lower SAM.

\subsection{Ultris-RPi real-device evaluation}

\begin{table*}[!t]
	\centering
	\resizebox{\textwidth}{!}{%
		\begin{tabular}{cclccccc}
			\toprule
			& SF & Method & mPSNR $\uparrow$ & RMSE $\downarrow$ & mSSIM $\uparrow$ & SAM [deg] $\downarrow$ & Time [s] $\downarrow$ \\
			\midrule
			\multirow{11}{*}{\rotatebox{90}{Real~\cite{lai2025hsifn}}}
			& \multirow{6}{*}{\rotatebox{90}{$\times4$}}
			& HSIFN~\cite{lai2025hsifn}
			& $31.9786{\pm}2.7543$ & $0.0275{\pm}0.0109$ & $0.8436{\pm}0.0369$ & $3.7786{\pm}1.0926$ & $1.2328{\pm}0.0533$ \\
			&& ESSAformer~\cite{zhang2023essaformer}
			& $32.7880{\pm}3.2412$ & $0.0294{\pm}0.0131$ & $0.8968{\pm}0.0479$ & $4.8850{\pm}2.1632$ & $0.4776{\pm}0.0109$ \\
			&& SSC-HSR~\cite{zhang2025sschsr}
			& $34.8129{\pm}3.8667$ & $0.0237{\pm}0.0121$ & $0.9224{\pm}0.0393$ & $2.2074{\pm}0.5635$ & $0.4764{\pm}0.0203$ \\
			\cmidrule(lr){3-8}
			&& CMF-GSA
			& $34.2240{\pm}3.7267$ & $0.0248{\pm}0.0113$ & $0.9234{\pm}0.0324$ & $2.3894{\pm}0.5173$ & $\mathbf{0.0461{\pm}0.0056}$ \\
			&& CMF-GSA-w
			& $34.1732{\pm}3.6950$ & $0.0250{\pm}0.0114$ & $0.9281{\pm}0.0294$ & $2.3077{\pm}0.5429$ & $0.0922{\pm}0.0063$ \\
			&& \textbf{CMF-GSA-wg}
			& $\mathbf{37.3442{\pm}3.4779}$ & $\mathbf{0.0168{\pm}0.0074}$ & $\mathbf{0.9434{\pm}0.0211}$ & $\mathbf{2.1697{\pm}0.5515}$ & $0.0956{\pm}0.0075$ \\
			\cmidrule(){2-8}
			& \multirow{5}{*}{\rotatebox{90}{$\times8$}}
			& HSIFN~\cite{lai2025hsifn}
			& $28.6940{\pm}2.9490$ & $0.0423{\pm}0.0167$ & $0.7609{\pm}0.0652$ & $4.1891{\pm}1.2229$ & $1.2325{\pm}0.0579$ \\
			&& SSC-HSR~\cite{zhang2025sschsr}
			& $28.7510{\pm}3.8835$ & $0.0480{\pm}0.0221$ & $0.7829{\pm}0.1079$ & $3.3277{\pm}1.0593$ & $0.4753{\pm}0.0194$ \\
			\cmidrule(lr){3-8}
			&& CMF-GSA
			& $32.8469{\pm}3.6326$ & $0.0283{\pm}0.0116$ & $0.8978{\pm}0.0465$ & $3.2291{\pm}0.8154$ & $\mathbf{0.0486{\pm}0.0011}$ \\
			&& CMF-GSA-w
			& $32.7655{\pm}3.6501$ & $0.0287{\pm}0.0119$ & $0.9001{\pm}0.0462$ & $3.1147{\pm}0.8751$ & $0.0911{\pm}0.0034$ \\
			&& \textbf{CMF-GSA-wg}
			& $\mathbf{33.7938{\pm}3.8149}$ & $\mathbf{0.0257{\pm}0.0115}$ & $\mathbf{0.9039{\pm}0.0435}$ & $\mathbf{2.9398{\pm}0.8248}$ & $0.0939{\pm}0.0025$ \\
			\bottomrule
		\end{tabular}%
	}%
	\caption{Quantitative comparison on the Real test set. Values are mean $\pm$ standard deviation over 10 test frames. CMF-GSA-w uses confidence only in the weighted regression, while CMF-GSA-wg additionally gates the final RGB-guided detail residual.}
	\label{tab:real_main_results}
\end{table*}

\begin{table}[!t]
	\centering
	\footnotesize
	\resizebox{0.75\linewidth}{!}{%
		\begin{tabular}{lcccc}
			\toprule
			Method
			& Total [M]
			& Reg. [M]
			& SR/fusion [M]
			& Peak mem. [GiB] $\downarrow$ \\
			\midrule
			HSIFN~\cite{lai2025hsifn}
			& 21.01 & 18.75 & 2.26 & 7.80 \\
			
			ESSAformer~\cite{zhang2023essaformer}
			& 11.64 & -- & 11.64 & 3.97 \\
			
			SSC-HSR~\cite{zhang2025sschsr}
			& 27.45 & 5.26 & 22.19 & 2.63 \\
			
			\cmidrule(lr){1-5}
			
			\textbf{CMF-GSA-wg}
			& 39.76 & 39.76 & $\mathbf{0}$ & $\mathbf{0.69}$ \\
			\bottomrule
		\end{tabular}%
	}%
	\caption{Learned parameter counts and peak allocated GPU memory. Parameter counts are decomposed into registration and SR/fusion components.}
	\label{tab:efficiency}
\end{table}

\begin{table*}[!t]
	\centering
	\resizebox{\textwidth}{!}{%
		\begin{tabular}{cclccccc}
			\toprule
			& SF & Energy variant & mPSNR $\uparrow$ & RMSE $\downarrow$ & mSSIM $\uparrow$ & SAM [deg] $\downarrow$ & Time [s] $\downarrow$ \\
			\midrule
			\multirow{8}{*}{\rotatebox{90}{Real~\cite{lai2025hsifn}}}
			& \multirow{4}{*}{$\times4$}
			& \textbf{Full gate}
			& $\mathbf{37.3442{\pm}3.4779}$ & $\mathbf{0.0168{\pm}0.0074}$ & $\mathbf{0.9434{\pm}0.0211}$ & $\mathbf{2.1697{\pm}0.5515}$ & $0.0956{\pm}0.0075$ \\
			
			&& w/o cycle cue $(\gamma=0)$
			& $37.3099{\pm}3.4378$ & $\mathbf{0.0168{\pm}0.0073}$ & $0.9428{\pm}0.0212$ & $2.1746{\pm}0.5510$ & $0.0504{\pm}0.0009$ \\
			
			&& w/o RGB res. $(\alpha=0)$
			& $35.3151{\pm}3.7098$ & $0.0215{\pm}0.0099$ & $0.9312{\pm}0.0310$ & $2.3445{\pm}0.5246$ & $0.0932{\pm}0.0038$ \\
			
			&& w/o energy $(\alpha=\gamma=0)$
			& $34.1760{\pm}3.7427$ & $0.0250{\pm}0.0115$ & $0.9257{\pm}0.0316$ & $2.3808{\pm}0.5071$ & $\mathbf{0.0502{\pm}0.0010}$ \\
			
			\cmidrule(lr){2-8}
			
			& \multirow{4}{*}{$\times8$}
			& \textbf{Full gate}
			& $33.7938{\pm}3.8149$ & $0.0257{\pm}0.0115$ & $\mathbf{0.9039{\pm}0.0435}$ & $\mathbf{2.9398{\pm}0.8248}$ & $0.0939{\pm}0.0025$ \\
			
			&& w/o cycle cue $(\gamma=0)$
			& $\mathbf{33.8194{\pm}3.7854}$ & $\mathbf{0.0256{\pm}0.0114}$ & $0.9036{\pm}0.0438$ & $2.9572{\pm}0.8343$ & $0.0509{\pm}0.0009$ \\
			
			&& w/o RGB res. $(\alpha=0)$
			& $33.2308{\pm}3.6130$ & $0.0270{\pm}0.0108$ & $0.9032{\pm}0.0439$ & $3.1343{\pm}0.8089$ & $0.0931{\pm}0.0026$ \\
			
			&& w/o energy $(\alpha=\gamma=0)$
			& $32.7810{\pm}3.6606$ & $0.0286{\pm}0.0118$ & $0.8988{\pm}0.0465$ & $3.2156{\pm}0.7978$ & $\mathbf{0.0506{\pm}0.0012}$ \\
			\bottomrule
		\end{tabular}%
	}%
	\caption{Confidence-cue ablation on the Real test set.
		Values are mean $\pm$ standard deviation over 10 test frames.
		The RGB residual is the dominant confidence cue. Removing cycle consistency gives reconstruction metrics nearly tied to the full model while roughly halving runtime by avoiding the backward flow.}%
	\label{tab:confidence_ablation}%
\end{table*}

\begin{table*}[!t]
	\centering
	\resizebox{\textwidth}{!}{%
		\begin{tabular}{ccclccccc}
			\toprule
			& SF & Bands & Method
			& mPSNR $\uparrow$
			& RMSE $\downarrow$
			& mSSIM $\uparrow$
			& SAM [deg] $\downarrow$
			& Time [s] $\downarrow$ \\
			\midrule
			
			\multirow[c]{3}{*}{%
				\rotatebox[origin=c]{90}{\footnotesize Ultris-RPi}%
			}
			& \multirow{3}{*}{$\times4$}
			& 31
			& SSC-HSR~\cite{zhang2025sschsr}
			& $29.3323{\pm}0.0925$
			& $0.0342{\pm}0.0004$
			& $0.8637{\pm}0.0017$
			& $1.8125{\pm}0.0100$
			& $0.2306{\pm}0.0179$ \\
			
			&& 31
			& \textbf{CMF-GSA-wg}
			& $\mathbf{29.8677{\pm}0.0901}$
			& $\mathbf{0.0322{\pm}0.0003}$
			& $\mathbf{0.9087{\pm}0.0010}$
			& $\mathbf{1.7713{\pm}0.0132}$
			& $\mathbf{0.0873{\pm}0.0064}$ \\
			
			\cmidrule(lr){3-9}
			
			&& 51
			& CMF-GSA-wg
			& $29.7565{\pm}0.1034$
			& $0.0329{\pm}0.0004$
			& $0.8909{\pm}0.0012$
			& $1.7122{\pm}0.0081$
			& $0.0957{\pm}0.0086$ \\
			
			\bottomrule
		\end{tabular}%
	}%
	\caption{Reduced-resolution quantitative evaluation on the 34-frame static Ultris-RPi sequence at $\times4$. Values are mean $\pm$ standard deviation over 34 frames. The 31-band VIS rows provide a direct comparison between SSC-HSR and CMF-GSA-wg, while the 51-band row evaluates CMF-GSA-wg on the full VIS-NIR spectral support.}%
	\label{tab:ultris_reduced}%
\end{table*}

\begin{figure*}[!t]
	\centering
	\begin{minipage}[t]{0.24\textwidth}
		\centering
		\includegraphics[width=\linewidth]{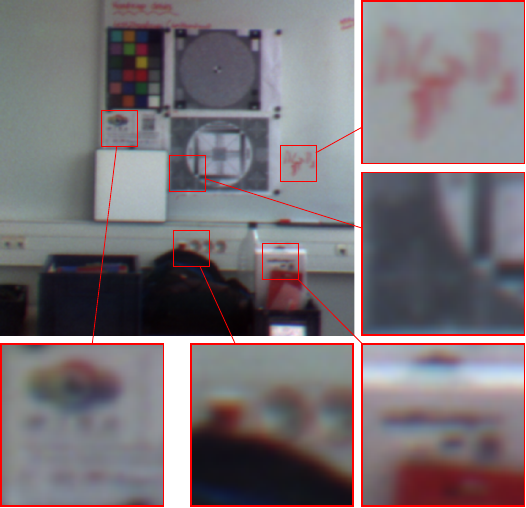}\\
		{\small\textbf{(a)} LR HSI}
	\end{minipage}%
	\hfill
	\begin{minipage}[t]{0.24\textwidth}
		\centering
		\includegraphics[width=\linewidth]{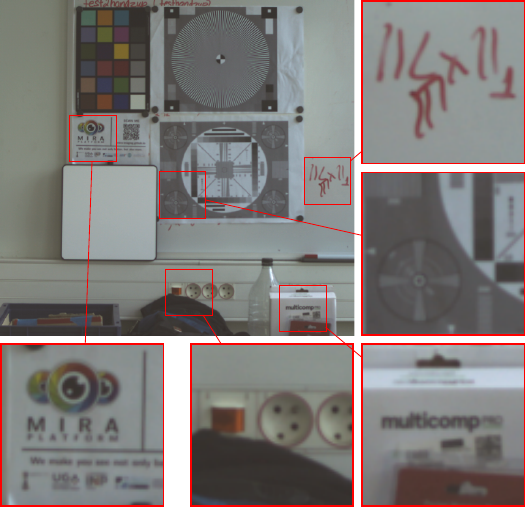}\\
		{\small\textbf{(b)} RGB-guide}
	\end{minipage}
	\hfill
	\begin{minipage}[t]{0.24\textwidth}
		\centering
		\includegraphics[width=\linewidth]{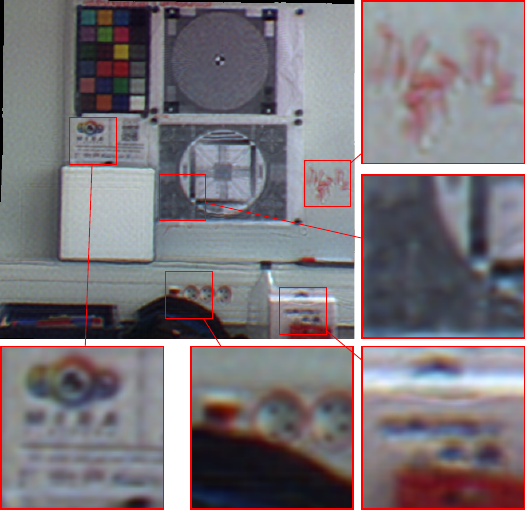}\\
		{\small\textbf{(c)} SSC-HSR~\cite{zhang2025sschsr}}
	\end{minipage}
	\hfill
	\begin{minipage}[t]{0.24\textwidth}
		\centering
		\includegraphics[width=\linewidth]{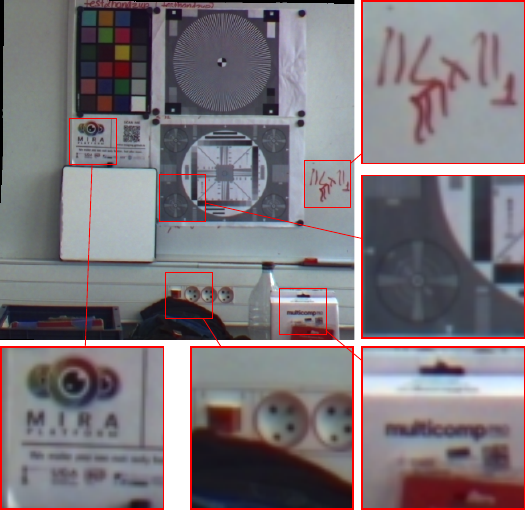}\\
		{\small\textbf{(d)} Ours, 51-bands}
	\end{minipage}
	\begin{minipage}[t]{0.24\textwidth}
		\centering
		\includegraphics[width=\linewidth]{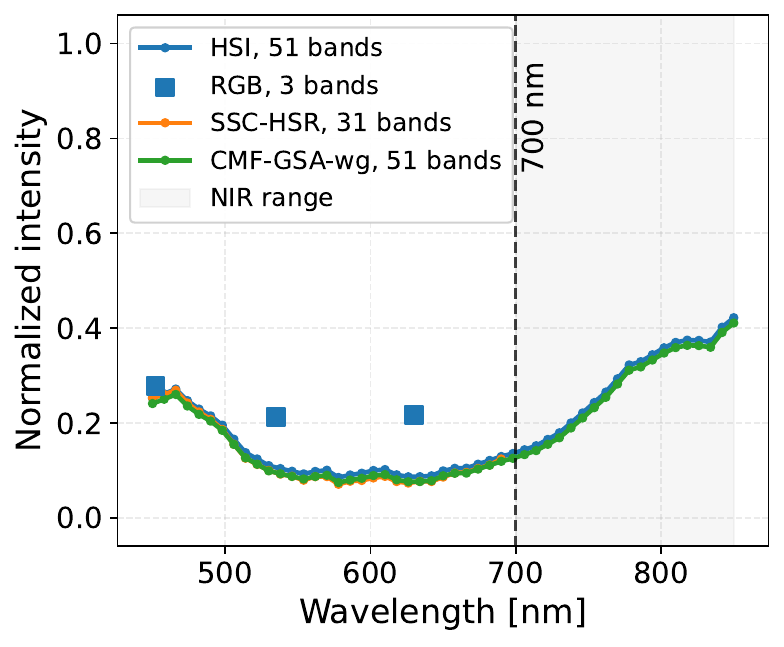}\\
		{\small\textbf{(e)} Blue square}
	\end{minipage}
	\hfill
	\begin{minipage}[t]{0.24\textwidth}
		\centering
		\includegraphics[width=\linewidth]{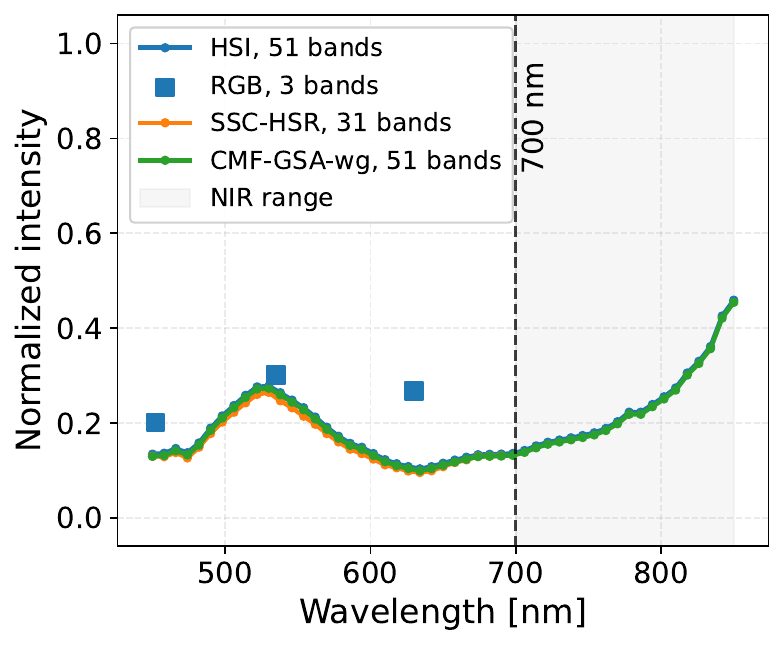}\\
		{\small\textbf{(f)} Green square}
	\end{minipage}
	\hfill
	\begin{minipage}[t]{0.24\textwidth}
		\centering
		\includegraphics[width=\linewidth]{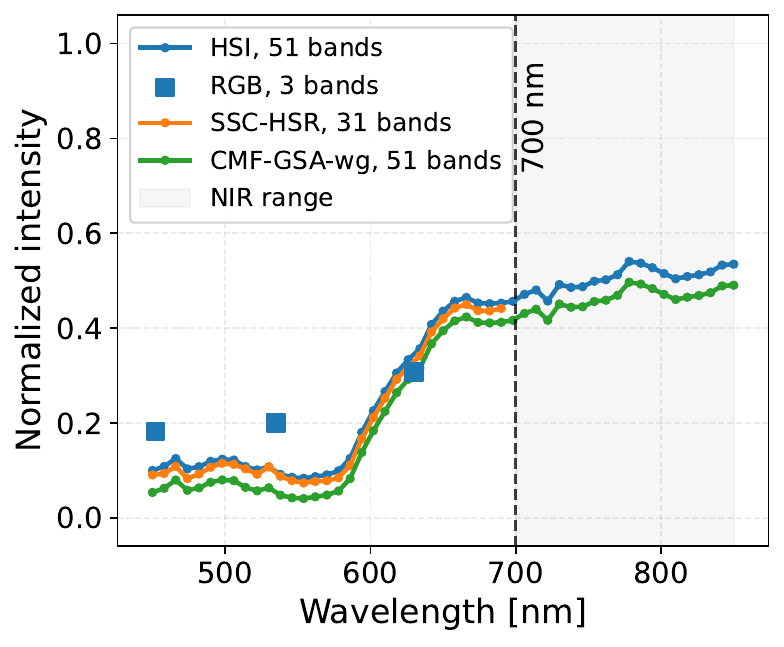}\\
		{\small\textbf{(g)} Red square}
	\end{minipage}
	\hfill
	\begin{minipage}[t]{0.24\textwidth}
		\centering
		\includegraphics[width=\linewidth]{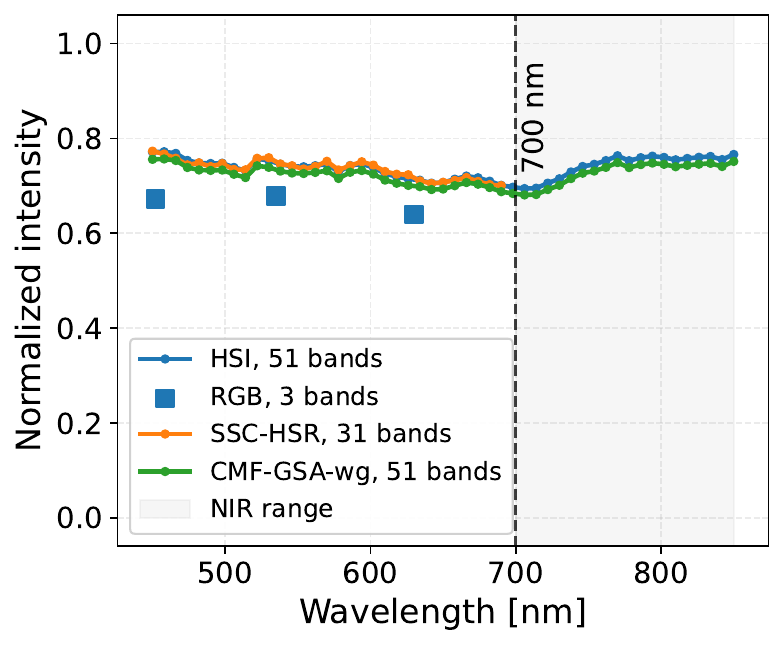}\\
		{\small\textbf{(h)} White reference}
	\end{minipage}%
	\caption{%
		Qualitative real-device evaluation on an Ultris-RPi frame.
		The first row shows visible-domain previews of the observed HSI, the observed RGB image, SSC-HSR, and the proposed CMF-GSA-wg reconstruction.
		The second row compares spectra at selected image locations. The observed HSI and CMF-GSA-wg curves cover the full 51-band VIS--NIR range, SSC-HSR is restricted to the 31-band VIS subset, and the RGB observation provides only three broadband measurements.%
	}%
	\label{fig:ultris_rpi_qualitative}%
\end{figure*}

\paragraph{Reduced-resolution evaluation.} \tablename~\ref{tab:ultris_reduced} reports the 34-frame reduced-resolution
evaluation. On the common 31-band VIS support, CMF-GSA-wg outperforms
SSC-HSR across all four reconstruction metrics, while the 51-band result
demonstrates direct deployment over the full VIS-NIR support without
retraining.

\paragraph{Native-resolution evaluation.} At native resolution, \figurename~\ref{fig:ultris_rpi_qualitative} provides the complementary qualitative evaluation. CMF-GSA-wg operates directly on the full 51-band VIS-NIR cube, whereas SSC-HSR remains restricted to its 31-band VIS setting.
As shown in the zoomed crops in \figurename~\ref{fig:ultris_rpi_qualitative}(c)
and (d), SSC-HSR produces clear local spatial artifacts under this real cross-sensor setting, while CMF-GSA-wg transfers the corresponding RGB details without retraining.
Since no clean HR HSI reference is available at this resolution, these results are interpreted qualitatively rather than as reference-based reconstruction metrics.

\section{Discussion}

The experiments support the proposed lightweight model-based alternative to learned unaligned RGB-guided HSR.
On Real, the main reconstruction gain comes from the final confidence gate rather than confidence-weighted regression alone. WLS modifies the global RGB-HSI spectral calibration, whereas the gate locally suppresses unreliable high-frequency RGB residuals.
The RGB-proxy residual is the dominant confidence cue on Real, while cycle consistency is mainly auxiliary, since forward-backward disagreement can also arise from the cross-resolution asymmetry between a smooth HSI proxy and a sharper RGB image.

The Ultris-RPi experiments highlight the deployment flexibility of the model-based formulation beyond the controlled Real protocol.
The 34-frame reduced-resolution evaluation validates the method on genuine cross-sensor data, while the native-resolution experiment demonstrates direct operation on the full 51-band VIS-NIR support without retraining, whereas the learned baseline remains restricted to its 31-band VIS setting.

The approach has limitations. First, the confidence is an energy-based reliability score, not a calibrated posterior probability. Second, if the HSI-to-RGB proxy is spectrally inaccurate or poorly radiometrically matched to the RGB camera, the RGB residual can confuse spectral-response mismatch with geometric misregistration. On real-device data, replacing the raw residual with a ZNCC-based structure cue reduces sensitivity to radiometry, but remains a heuristic reliability cue rather than a calibrated geometric uncertainty. Third, while confidence gating reduces the impact of local flow errors, it misses RGB details where no reliable correspondence exists, leading to local low-resolution fallbacks. That said, stronger HSI-only priors, such as spatial-spectral regularization~\cite{condat2023proximal} or HSI single-image super-resolution fallback models, may improve large unreliable regions.
Future work should include real-device RGB-HSI benchmarks with clean HR HSI references.

\section{Conclusion}

We presented a lightweight interpretable framework for RGB-guided hyperspectral super-resolution under real cross-resolution misalignment.
The method combines cross-modal flow and GSA with an explicit reliability map used for both confidence-weighted spectral regression and gated detail injection.
On Real, the proposed method improves over learned unaligned fusion baselines while remaining substantially faster, with the final gate providing the main reconstruction gain.
On a real RGB-HSI setup, the 34-frame Ultris-RPi evaluation further validates the method under real cross-sensor differences, while native-resolution results demonstrate deployment on a full 51-band VIS-NIR cube without retraining.

\section*{Acknowledgements}

This work was supported by the Institut Carnot Logiciels et Systèmes Intelligents (Carnot LSI) and by the French National Research Agency (ANR) under the France 2030 programme with reference ANR-23-IACL-0006.
The computations were performed using the GRICAD infrastructure (GRANT CPER07\_13 CIRA, ANR10 LABX56, ANR-10-EQPX-29-01).
Data acquisitions were carried out with the support of the
Multi-camera Imaging Research and Acquisition (MIRA) Platform  (\url{https://mira-imaging.github.io/})
of GIPSA-lab, which provided the camera setup and the acquisition and research resources used in this work.
The MIRA Platform is developed with the financial support of the Institut Carnot Logiciels et Systèmes Intelligents (Carnot LSI).
The authors thank Dr. Amaury Negre for his contributions to the ROS-based acquisition software and to the acquisition of the Ultris-RPi sequence.

	\bibliography{references_draft}	

\end{document}